\documentclass[conference]{IEEEtran}
\IEEEoverridecommandlockouts
\usepackage{cite}
\usepackage{amsmath,amssymb,amsfonts}
\usepackage{graphicx}
\usepackage{textcomp}
\usepackage{xcolor}

\usepackage{multirow}
\usepackage{booktabs}
\usepackage{tabularx}
\usepackage{lipsum}
\usepackage{enumitem}
\usepackage[hidelinks]{hyperref}
\usepackage{subcaption}
\usepackage{algorithm}
\usepackage[noend]{algpseudocode}
\usepackage{bm}
\usepackage{marginnote}
\usepackage{xcolor}

\usepackage{colortbl}
\definecolor{Gray}{gray}{0.9}

\newcommand{\mypar}[1]{\vspace{0.5pt}\noindent\textbf{#1.}}
\newcommand{\mypartwo}[1]{\vspace{0.5pt}\noindent\textit{#1.}}

\begin{document}

\title{AgentSimulator: An Agent-based Approach for Data-driven Business Process Simulation\\
}

\makeatletter
\newcommand{\linebreakand}{%
  \end{@IEEEauthorhalign}
  \hfill\mbox{}\par
  \mbox{}\hfill\begin{@IEEEauthorhalign}
}
\makeatother


\author{
\IEEEauthorblockN{
Lukas Kirchdorfer\IEEEauthorrefmark{1}\IEEEauthorrefmark{2}, 
Robert Blümel\IEEEauthorrefmark{1}, 
Timotheus Kampik\IEEEauthorrefmark{1}, 
Han van der Aa\IEEEauthorrefmark{3} and
Heiner Stuckenschmidt\IEEEauthorrefmark{2}}
\IEEEauthorblockA{\IEEEauthorrefmark{1}SAP Signavio, Walldorf, Germany \\ Emails: lukas.kirchdorfer@sap.com, robert.bluemel@sap.com, timotheus.kampik@sap.com} 
\IEEEauthorblockA{\IEEEauthorrefmark{2}University of Mannheim, Mannheim, Germany\\ Email: heiner.stuckenschmidt@uni-mannheim.de}
\IEEEauthorblockA{\IEEEauthorrefmark{3}University of Vienna, Vienna, Austria \\ Email: han.van.der.aa@univie.ac.at} }

\maketitle

\begin{abstract}
Business process simulation (BPS) is a versatile technique for estimating process performance across various scenarios. Traditionally, BPS approaches employ a control-flow-first perspective by enriching a process model with simulation parameters. Although such approaches can mimic the behavior of centrally orchestrated processes, such as those supported by workflow systems, current control-flow-first approaches cannot faithfully capture the dynamics of real-world processes that involve distinct resource behavior and decentralized decision-making. Recognizing this issue, this paper introduces \textit{AgentSimulator}, a resource-first BPS approach that discovers a multi-agent system from an event log, modeling distinct resource behaviors and interaction patterns to simulate the underlying process. Our experiments show that AgentSimulator achieves state-of-the-art simulation accuracy with significantly lower computation times than existing approaches while providing high interpretability and adaptability to different types of process-execution scenarios.
\end{abstract}


\section{Introduction}

Business process simulation (BPS) is a widely used technique to estimate the impact of changes to a process with respect to key performance indicators, such as cycle time, resource utilization, or waiting time for a given activity---a practice known as counterfactual reasoning or ``what-if'' analysis~\cite{FundamentalsOfBPM}. BPS has the potential to drastically reduce the risks of process change and facilitate process improvement with tangible outcomes, as decision makers can compare process designs without already having to implement the changes. Nevertheless, the effectiveness of BPS relies heavily on the availability of a simulation model that precisely mirrors the dynamics of a given process across dimensions such as control-flow, time, and resource behavior. Manually constructing these simulation models is time-consuming and error-prone due to several pitfalls \cite{Aalst15}. 
Therefore, various approaches have been developed for the automated discovery of process simulation models based on historical execution data contained in event logs~\cite{RozinatMSA09,Camargo_2020,meneghello_RIMS,camargo_DSIM,KhodyrevP14}.
Most commonly, such data-driven simulation approaches discover a process model---capturing the control flow of an entire process---and subsequently augment this model with simulation parameters, such as arrival rates, resources, and processing times.

In this paper, we argue that there are settings in which such \textit{control-flow-first} simulation models cannot provide a faithful representation of the dynamics of real-world processes, leading to simulation inaccuracies.
This particularly applies to processes 
involving distinct resource behavior or decentralized decision-making.
Although certain processes are indeed centrally orchestrated, such as those supported by workflow systems~\cite{FundamentalsOfBPM}, 
other processes provide higher operational flexibility to the actors involved.
In such settings, each actor in a process performs their part from their own perspective and, to some extent, in their own manner, i.e., they receive a case from a co-worker, conduct one or more tasks they deem necessary, and pass on the case to the next individual or system.
This can result in processes where specific traits or preferences of actors can influence a case's execution. Such characteristics are difficult for existing control-flow-first approaches to capture.

Therefore, we use this paper to propose  \textit{AgentSimulator}, a resource-first approach for data-driven process simulation. By discovering a multi-agent system (MAS) from an event log, AgentSimulator can simulate the execution of a process through autonomous and interacting agents, corresponding to real-world actors and systems. This allows AgentSimulator to achieve various benefits in comparison to existing approaches for data-driven process simulation:
\begin{itemize}[noitemsep,topsep=0pt]
    \item our approach provides full flexibility over the behavior of individual resources involved in a process, allowing to capture differences in terms of control-flow behavior, interaction preferences, and capabilities;
    
    \item agent-based systems are highly interpretable and allow for a high degree of adaptability, which is crucial to performing what-if analyses, and thus provides a substantial advantage over black-box deep learning models;

    \item our approach requires significantly lower run times than previous approaches, making it more feasible to run a large number of simulations, which is necessary to achieve stable and confident results given the stochastic nature of simulation models; 

    \item and finally, our approach yields state-of-the-art simulation accuracy across a variety of event logs.
    
\end{itemize}

\noindent The remainder starts with a motivating scenario (\autoref{sec:motivation}), followed by the presentation of the AgentSimulator approach itself (\autoref{sec:our_approach}). Then,
\autoref{sec_evaluation} reports on evaluation experiments, highlighting the benefits of our approach. Finally, \autoref{sec:rel_work} discusses related work, before \autoref{sec:conclusion} concludes the paper.

\section{Motivation}
\label{sec:motivation}

This section illustrates the benefits of shifting simulation models from a control-flow-first to a resource-first perspective. 

For this illustration, we consider a simplified credit application process, 
for which a schematic visualization is shown in \autoref{fig:motivation_example}.
As depicted, the process starts when an application is received by the system, after which the applicant's credit history and income sources need to be checked by a clerk (in any order). Once both checks have been completed, the application is passed on to a credit officer, who assesses the application and notifies the applicant of the outcome.
As shown,
there are three clerks (Steve, Oliver, and Angela) and two credit officers (Maria and Patrick) involved in the process.

Even for such a simple scenario, we may observe various ways in which the involvement of specific actors in a case can influence its execution:

\begin{itemize}[noitemsep,topsep=0pt]
    \item \textit{Process performance.} The execution time of an activity may depend on the employee who performs it. For example, Steve (a less experienced, Junior Clerk) might need 45 minutes to check the credit history and the income sources, respectively. For the same activities, the Senior Clerks Oliver and Angela only need 15 to 20 minutes. Considering these performance differences between resources is critical for accurate simulation, as has been recently demonstrated~\cite{Lopez-PintadoD22}.

    \item \textit{Resource availability.} Employees involved in a process may have different availabilities, owing to factors such as part-time work and other duties.  Such considerations are particularly relevant in decentralized processes, where cases 
    may be handed over directly from employee to employee, rather than by a central workflow system that can assign a case to the next available person.  For example, if Angela hands over an application specifically to Patrick, rather than to the next available credit officer, this can lead to considerable delays in the case's execution if Patrick is not available for the next working days. Such irregularities should be reflected in a simulation model, calling for individual resource calendars, as again was recently demonstrated to positively impact accuracy~\cite{Lopez-PintadoD22}.

    \item \textit{Control-flow behavior.} Control-flow-first simulation models impose the assumption that the sequence of activities performed for a given case is independent of the actors involved in it. However, there can be various reasons why this is not the case. In our scenario, for example, we may observe that Angela always checks the credit history first, before checking the income sources, whereas other actors may alternate these orders. Furthermore, there may even be actor-specific rules that influence the possible sequences of a case. For example, it may be necessary that any application handled by a Junior Clerk (e.g., Steve) needs to go through an additional verification step performed by a Senior Clerk.
    Capturing such actor-dependent behavior is difficult and often generally omitted from control-flow-first simulation models, despite it having a considerable impact on process execution.

    \item \textit{Interaction patterns.} Finally, owing to a lack of central orchestration, 
    there may be specific interaction patterns among actors. For example, we may observe that 
    Steve always collaborates with Oliver, since they both work in Mannheim, while he has no contact with Angela in Hamburg. Such specific patterns influence the workload of individual resources, e.g., leading to an imbalance between Oliver and Angela, which is missed by general simulation models. Furthermore, as described above, specific interaction patterns may lead to additional delays when the availability of individual resources differs.
\end{itemize}

\noindent All of the above factors influence the execution of cases in a process. Therefore, process simulation models should reflect these as faithfully as possible in order to appropriately mimic the dynamics of a real-world process. The former two aspects have already been recognized and captured by control-flow-first models~\cite{Lopez-PintadoD22}. However, especially the latter two can be much more naturally incorporated by an agent-based simulation model, where the behavior of each resource is explicitly captured. This is achieved by our AgentSimulator approach, described next.

\begin{figure}
    \centering
    \includegraphics[trim = 0mm 0mm 0mm 0mm, clip, width=\columnwidth]{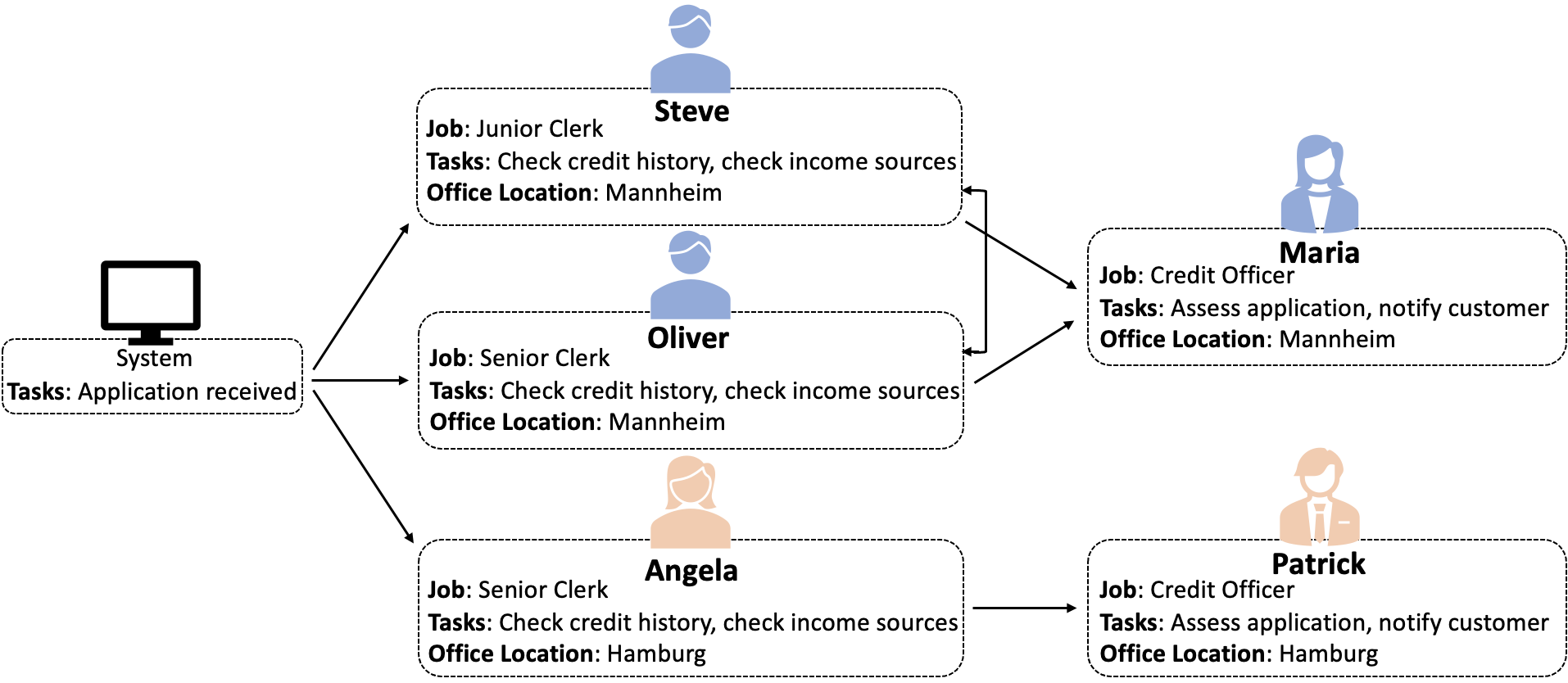}
    \caption{Sketch of the credit application process.}
    \label{fig:motivation_example}
\end{figure}

\section{Our Approach: AgentSimulator}
\label{sec:our_approach}

\begin{figure*}
    \centering
    \includegraphics[trim = 0mm 55mm 0mm 43mm, clip, width=1.5\columnwidth]{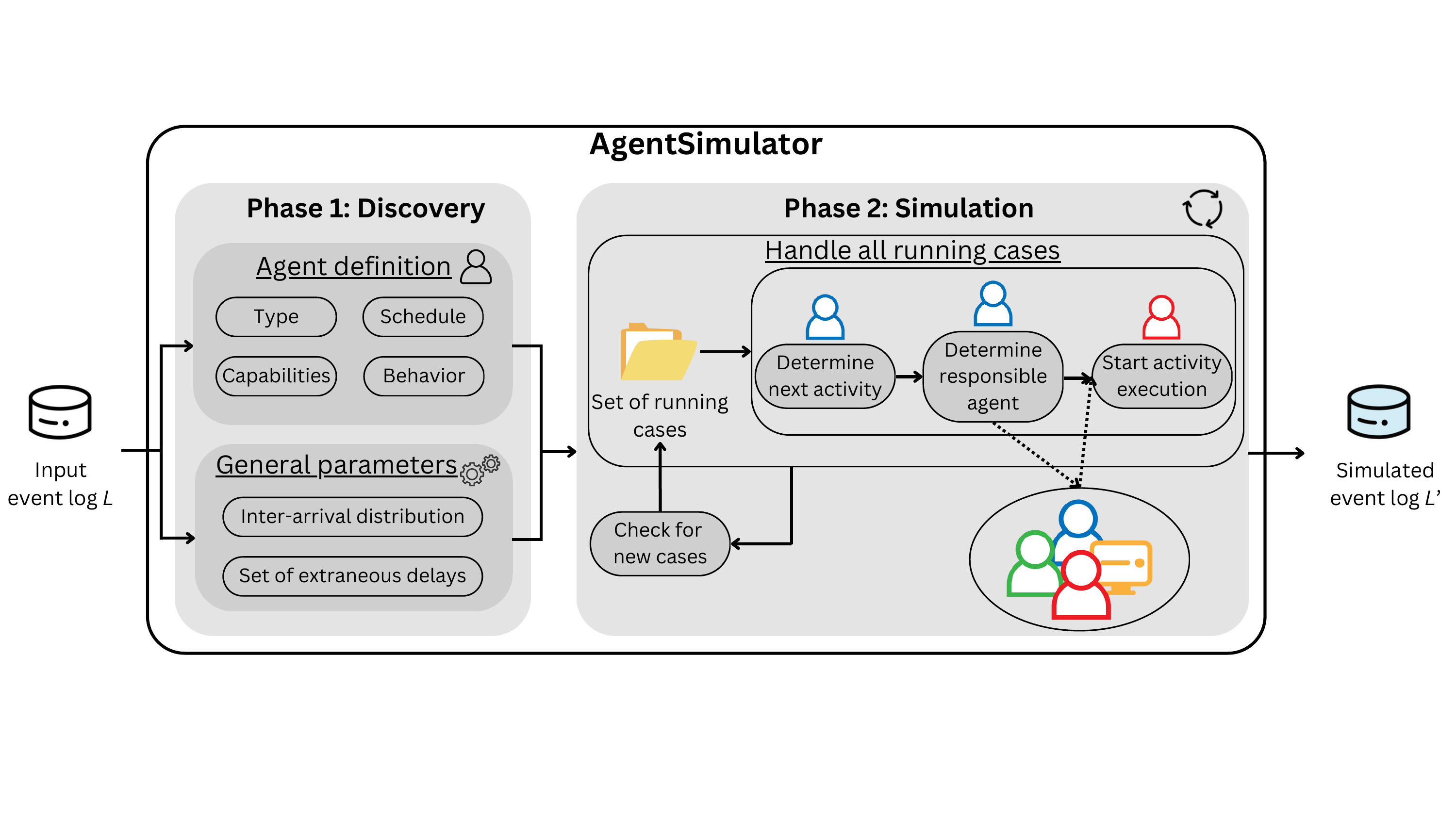}
    \caption{Overview of the end-to-end AgentSimulator approach.}
    \label{fig:asim_overview}
\end{figure*}

This section introduces AgentSimulator, our data-driven agent-based business process simulation approach, with a high-level overview in \autoref{fig:asim_overview}.
To adopt a resource-first perspective, AgentSimulator models agents in a MAS (see \autoref{sec:definitions}), which is (along with general simulation parameters) discovered from an event log in the discovery phase (see \autoref{sec:discovery}). Following the MAS discovery, we can simulate the process and generate a new event log (see \autoref{sec:simulation}). We detail our approach for both discovery and simulation, also discussing alternative design choices for different process types, highlighting AgentSimulator's adaptability.

\subsection{Definitions}
\label{sec:definitions}
In this section, we define event logs, which provide the input of our approach (and its simulated output), before providing the definition of agents and a MAS that our approach employs.

\mypar{Input} AgentSimulator takes as input an event log $L$, which we define as a finite multi-set of traces. A trace  $\sigma \in L$ is a finite sequence of events, $\langle e_1,...,e_n\rangle$, recording the execution of activities performed for a single case in an organizational process. Each event $e_i$ is a tuple $(act,ts_{start},ts_{end},res)$, with 
$act$ as the activity to which it corresponds, $ts_{start}$ and $ts_{end}$, respectively, as the start and end timestamps of the activity's execution, and $res$ as the resource that executed the activity. 
Note that we follow \cite{Lopez-PintadoD22} by representing each event with a start and end timestamp, which is required in simulation settings to consider activity durations, and ordering the events in a trace based on their start timestamp. 

In the remainder, we commonly use dot notation to refer to components of tuples, e.g., using $e_i.act$ as a shorthand to refer to the activity of an event $e_i$, whereas we use $ACT_L$ and $RES_L$ to, respectively, refer to the sets of activities and resources contained in the traces of event log $L$. 

\mypar{Multi-agent system} The notion of an agent is a fundamental abstraction in Artificial Intelligence (AI) \cite{RN2020}. Agents perceive their environment (including other agents) to reason about the perceptions and decide on an action, which is then executed against the environment. In a multi-agent system (MAS), agents can collaborate to achieve joint goals. In our work, we define a MAS and an agent as follows:

\mypar{Definition 1 (Multi-agent system)} We define a MAS for our AgentSimulator as a tuple $m = (A, p)$. Here, $A$ corresponds to the set of agents used to simulate a process, whereas $p$ is a tuple of general simulation parameters, reflecting aspects of the process's environment. Here, $p$ consists of a case inter-arrival distribution and a set of probability density functions (PDFs) over extraneous delays (more details in \autoref{sec:discovery}).

\mypar{Definition 2 (Agent)} An agent in a business process environment is an autonomous entity representing a real-world actor or system. We define an agent $a \in A$ that acts in the MAS $m$ as a tuple, denoted $a = (t, s, c, b)$, where:

\begin{enumerate}
    \item $t$ refers to the agent's \textbf{type}. Agent types are used to indicate agents with similar characteristics, e.g., in terms of the activities they can perform. We use $T$ to refer to the set of types in $A$

    \item $s$ refers to a \textbf{schedule} (a set of intervals) during which the agent is available to perform activities. The simulation needs to capture the distinct availabilities of agents (e.g., full-time vs. part-time) to faithfully reflect the temporal dimension of the process.

    \item $c$ refers to \textbf{capabilities} of an agent, denoted as a tuple $c = (\textsc{Alloc}, PT)$, where:
    \begin{itemize}
        \item $\textsc{Alloc} \subseteq ACT_L$ is the set of activities that $a$ can execute, i.e., that can be allocated to agent $a$.

        \item ${PT}$ refers to a set of PDFs, where each $f_{pt}(act) \in PT$ captures a distribution over processing times $pt \in [0,\infty)$ for an activity $act \in \textsc{Alloc}$. Defining $PT$ per agent allows us to differentiate process performance across resources.

    \end{itemize}

    \item $b$ refers to the \textbf{behavior} of an agent, capturing how agent $a$ hands a case over to continue its execution after finishing an activity.   What $b$ exactly entails depends on the kind of process being simulated. In a centrally orchestrated process, determining the next activity is handled centrally, whereas in a more autonomous process, agents directly hand over cases to others. Therefore, we provide further details and exact definitions in the following sections.

\end{enumerate}

\subsection{Phase 1: MAS Discovery}
\label{sec:discovery}
In this section, we describe how AgentSimulator discovers the agents $A$ and general simulation parameters $p$ from an event log $L$ to define the MAS $m$. In the realm of business processes, our generic definition of an agent offers numerous implementation possibilities. Thus, besides describing our discovery approach, we also discuss alternatives. The flexibility of agent systems is a key advantage of our AgentSimulator, enabling adaptation to diverse collaborative work dynamics. 

\mypar{Agent instantiation} In business process environments, resources are the active entities performing actions. Therefore, we instantiate one agent $a \in A$ for each resource $res \in RES_L$. For our motivating example, we thus instantiate six agents (five human actors and \textit{System}).  If certain events lack resource information, we additionally generate a dummy agent for each distinct activity that lacks such information. We noticed that events without resource information often refer to either instantaneous activities or system resources. First, explicitly modeling these as agents ensures that any activity can be associated with at least one agent, which is required for the simulation. Second, it allows discovering distinct behaviors for these agents to simulate their activities appropriately, e.g., to consider that performing instantaneous activities does not require waiting for a corresponding agent.

\mypar{Agent type} 
To assign a type to each agent, we use the algorithm from \cite{Song2008} that clusters agents into types based on the similarity of their executed activities. This algorithm is also used in \cite{Camargo_2020} to group simulated resources.

\mypar{Schedule} 
We discover a schedule for each agent $a \in A$ by using the same algorithm as proposed in \cite{Lopez-PintadoD22}. Note that discovering the schedule per agent type is also a possible design choice that our approach allows for. Furthermore, the more recent approach to discovering probabilistic instead of crisp calendars \cite{LopezPintado_Prob_Calendars} could also be integrated.

\mypar{Capabilities} The capabilities $a.c$ of an agent $a$ consist of a set of activities and a set of PDFs over processing times. 

\mypartwo{Set of activities} We discover the set of activities $a.c.\textsc{Alloc}$ that agent $a$ can execute by checking which activities from $ACT_L$ (the resource corresponding to) $a$ has performed in $L$. 

\mypartwo{Processing times} The set of PDFs over processing times $a.c.PT$ contains one distribution for each activity that reflects how long agent $a$ requires to execute each of these activities. Following \cite{Camargo_2020}, we fit various distributions to the activity durations. Subsequently, we select the distribution with the lowest error measured by the Wasserstein distance. The set of distributions includes the following ones: Exponential, Gamma, Normal, Uniform, Log-Normal, and a fixed value to capture activities with fixed durations. However, instead of discovering one PDF per activity, we discover one PDF for each combination of agent and activity, denoted as a mapping $ACT_L \times A \rightarrow PTT$, where $PTT = \bigcup_{a \in A} a.c.PT$. This ensures more realistic activity processing times, as, for example, a junior employee typically requires more time to perform the same activity than a senior employee.

\noindent The properties type, schedule, and capabilities are aspects commonly considered in non-agent-based simulation approaches as well, which is why we largely follow existing works for their discovery. However, the distinctive agent behaviors serve as the main ingredient of our AgentSimulator.

\mypar{Behavior} The behavior of an agent $a.b$ describes its activity transition and agent handover patterns. Therefore, the behavior of an agent determines (i) which activity is the next in an ongoing case and (ii) who performs this activity. The behavior can be learned in many different ways, leading to numerous design options for the architecture of the MAS. Below, we outline two distinct behavior discovery approaches: one suitable for capturing processes involving orchestrated handovers and one for processes with autonomous handovers.

\mypartwo{Orchestrated handovers} 
     We use this first configuration to mimic processes that are centrally orchestrated, such as those supported by a workflow or business process management system~\cite{FundamentalsOfBPM}.
     In such processes, the execution of a case is centrally guided. 
     Therefore, this kind of architecture only requires discovering global execution patterns, which are independent of specific agents.
     We achieve this by learning activity transition probabilities at the log level.
     
     Specifically, we compute the frequentist probability of transitioning to an activity given the activity prefix of an ongoing case, where a prefix 
     $\sigma_{\text{prefix}} = \langle e_1, e_2, ..., e_k \rangle$ is the sequence of events from the beginning of $\sigma$ up to event $e_k$, with $\sigma_\text{prefix}^{act}$ being the corresponding activity sequence.
     This transition probability $P(act | \sigma_\text{prefix}^{act})$ is computed by determining how often each possible activity prefix $\sigma_\text{prefix}^{act}$ in log $L$ is followed by each activity $act \in ACT_L$. This is achieved by dividing the number of times the specific transition from $\sigma_\text{prefix}^{act}$ to $act$ happens by the number of times the activity prefix occurs in the log. Note that, in case an activity prefix $\sigma_\text{prefix}^{act}$ has not been observed in the log, we iteratively remove the first activity of the prefix until we reach a subsequence that has been observed in $L$, e.g., if we have not seen $\langle a, b, c, d \rangle$, we next check for occurrences of $\langle b, c, d \rangle$. To transition to the end of a case, we introduce a placeholder end event following the last activity in each case. Furthermore, without loss of generality, transition probabilities may also be computed using other methods, such as through next-activity prediction techniques~\cite{camargo_DGEN, TaxVRD17}.

    \mypartwo{Autonomous handovers} As argued in \autoref{sec:motivation}, many processes provide higher flexibility and decision power to the agents involved. Therefore, our AgentSimulator can also be used to discover decentralized MASs     
    in which agents themselves determine the next activity for a process instance and which agent should perform it. 
    In this case, activity transition probabilities need to be learned locally instead of globally.
    The locality can either be given by the individual agent or the agent type, i.e., activity and agent transition probabilities can either be unique for each agent or generalized across all agents of the same type $t \in T$. The latter can make sense if only limited data is available per agent. 

    To compute transition probabilities from a given activity prefix $\sigma_\text{prefix}^{act}$ to an activity $act \in ACT_L$ individually for each agent $a \in A$, we can extend the computation described for orchestrated handovers and make it agent-depending. Thus, $P(act | \sigma_\text{prefix}^{act},a)$ is computed by counting for each agent $a \in A$ how often there happened to be a transition from $\sigma_\text{prefix}^{act}$ to $act$ with the last activity of the prefix being performed by $a$,
    divided by the total number of occurrences of $\sigma_\text{prefix}^{act}$ with its last activity being executed by $a$. Consider Angela in \autoref{fig:motivation_example} who, given the prefix \textit{application received}, always first checks the credit history before the income sources, whereas Steve and Oliver have no fixed order as they sometimes collaborate.

    In addition to activity transitions, the autonomous handover architecture also accounts for distinct interaction patterns among agents. 
    Therefore, we compute the frequentist probability of handing over a case from one agent to another agent. 

    The conditional probability $P(a_i|a_j)$ of handing over a task from agent $a_j$ to $a_i$ with $i,j \in \{1,...,|A|\}$ is computed by counting all occurrences where one activity is performed by $a_j$ and the following activity is executed by $a_i$, divided by the total number of activities performed by $a_j$. Following our example, this results in $P(Maria|Angela)=0.0$ and $P(Patrick|Angela)=1.0$.

    There exist several other approaches to compute the handover probabilities. For instance, including the dependence of the specific activity in the conditional probability $P(a_i|a_j,act)$ could make the interaction patterns even more precise. Comparing different approaches for determining the interaction patterns can be tackled in future work.

    The combination of $P(act | \sigma_\text{prefix}^{act},a)$ and $P(a_i|a_j)$ defines agent behavior in an autonomous handover architecture and captures agent-specifics that influence the progress of a case.

\mypar{General simulation parameters} After having instantiated the set of agents $A$, we discover some general simulation parameters $m.p = (f_{iat}, D)$, where $f_{iat}$ is a PDF of case inter-arrival times and $D$ a set of PDFs over extraneous delays. These parameters are not specific to an agent-based simulation model but are also used in other BPS approaches \cite{Camargo_2020}.

\mypartwo{Inter-arrival times} Inter-arrival times denote the duration between the start of two consecutive cases. The PDF over inter-arrival times is used during simulation to sample new cases. For discovering the PDF over case inter-arrival times $f_{iat}$, we follow~\cite{Camargo_2020} who fit different distributions to the inter-arrival times and take the one with the smallest error.

\mypartwo{Extraneous delays} Because different instances of a process typically compete for limited resources and because resources do not always start an activity as soon as it can possibly be performed, processes are affected by waiting times. Extraneous delays are waiting times that are not caused by resource contention or unavailability (e.g., a resource waits for the customer to return a phone call). They need to be modeled explicitly. Following the algorithm in~\cite{chapela2024enhancing}, we discover a PDF over extraneous delays for each activity, resulting in a set of extraneous delay distributions~$D$.

\subsection{Phase 2: Simulation}
\label{sec:simulation}
This section details how the AgentSimulator approach uses the discovered MAS $m$ to simulate an event log $L'$, illustrated by the pseudo-code in Algorithm \ref{alg:simulation}. The simulation proceeds in discrete steps, each representing a time tick. At each step, we check if new cases arrive and for each running case in the system, we verify if it is waiting to be processed (e.g., due to completion of the previous activity) to instantiate the next activity. The simulation step ends after all cases have been checked. In more detail, one simulation step looks as follows:

\mypar{1) Check for new cases} Based on the discovered inter-arrival distribution $p.f_{iat}$, we first check if a new case arrives (Line~\ref{line:new_case}). If there are multiple cases in the system, their processing order is determined in a first-in-first-out manner. Note that when evaluating against a test log, we follow the convention of Simod \cite{Camargo_2020} and only simulate as many case arrivals as there are cases to be simulated. However, for regular simulations, we keep on letting cases arrive until we finish the simulation to avoid a cool-down phase, which is more realistic. 

\mypar{2) Handle all running cases} The cases are processed one-by-one (Line~\ref{line:for_each_case}) until all cases have been addressed:

\mypartwo{a) Determine next activity} First, we check if the given case $\sigma_\text{prefix}$ is waiting to be processed (Line~\ref{line:check_waiting}). If so, a new event $e$ is created with its corresponding activity $e.act$ being determined (Line~\ref{line:get_activity}) either globally, modeling orchestrated handovers based on the current activity prefix (i.e., a global entity orchestrates the control-flow execution), or locally, modeling autonomous handovers through the last active agent in the case. Note that a case's first activity is always determined globally as there is no prior agent.

\mypartwo{b) Determine the responsible agent} After a new activity was determined for the given case, it requires an agent for the execution. Based on the set of activities that an agent can execute $a.c.\textsc{Alloc}$, we identify possibly responsible agents $p\_a$ (Line~\ref{line:potential_agents}). To determine the specific agent (Line~\ref{line:get_agent}), we again differentiate between the two architectures:
\begin{itemize}[noitemsep,topsep=0pt]
        \item \textit{Orchestrated handovers.} In a MAS with orchestrated handovers, there are no agent interaction patterns modeled. Thus, having determined the set of possibly responsible agents, we order this set based on the agent availabilities and iteratively ask these agents to perform the next activity until one agent is available to execute it (which we call iterative task allocation). Thus, the first agent to be asked is the one that offers (based on its schedule $a.s$ and its current occupation) the earliest availability. An agent refuses to accept the activity if its estimated duration (based on $f_{pt}(act) \in a.c.PT$) collides with occupied or non-working time slots. 
        If none of the possible agents are available, the case cannot be processed for now and will be checked again in the next simulation step. Thus, the case receives waiting time due to contention (Line~\ref{line:waiting}).

        \item \textit{Autonomous handovers.} In a MAS with autonomous handovers, we use the computed agent handover probabilities $P(a_i|a_j)$ to determine the next agent. There are two design choices for their usage: (i) iterative task allocation and (ii) direct task assignment. 
        
        (i) Using iterative task allocation, we simulate (just as in the orchestrated handovers) that agent $a_j$ iteratively asks other agents to take the activity. However, the difference is that here the order of to-be-asked agents is determined by the agent handover probabilities that are transformed into a ranking, i.e., agent $a_i$ is asked first if it has the highest handover probability from $a_j$.

        (ii) Using direct task assignment (not represented in Algorithm \ref{alg:simulation}), current agent $a_j$ does not ask agents until one is available but assigns the task to an agent who then starts executing it as soon as it finds the time. This can be a common agent interaction pattern in real-life processes, for instance, if an employee forwards a case via e-mail to a colleague. 
        For this direct assignment, we sample the agent based on the agent handover probabilities. 

\end{itemize}

\noindent Note that a case's first agent is always determined as described for orchestrated handovers.

\mypartwo{c) Start activity execution} If an agent could be identified, it begins executing the assigned activity, with the end timestamp being determined by sampling from the agent-specific activity duration distribution $f_{pt}(act) \in a.c.PT$ and potentially adding extraneous delays using $f_{d}(act) \in p.D$ (Line~\ref{line:execute_act}).

\noindent After each case is processed, a simulation step is completed. A case exits the system once its final activity is executed (Line~\ref{line:finish_case}). 
  The entire simulation concludes when the specified number of cases has been processed and finished. The output of the simulation is an event log $L'$, where each event is a tuple $e = (act, ts_{start}, ts_{end}, a)$.

\begin{algorithm}[t]
\footnotesize
\caption{Simulation}
\hspace*{\algorithmicindent} \textbf{Input:} $A$: set of agents; $n$: number of to-be simulated cases \\
\hspace*{\algorithmicindent} \textbf{Input:} $f_{iat}, D$: general simulation parameters
\begin{algorithmic}[1]
\State $\Sigma \gets \left \{ \right \}$, $L' \gets \left \{ \right \}$ \Comment{init. set of running traces $\Sigma$ and output log $L'$}
\While{$len(L') \leq n$}
        \State $\Sigma \gets check\_for\_new\_cases(f_{iat}, \Sigma, evaluation=False)$ \label{line:new_case}
    \For {$\sigma_\text{prefix}$ in $\Sigma$} \label{line:for_each_case}
        \If{$is\_waiting(\sigma_\text{prefix})$} \label{line:check_waiting}
            \State $e \gets ()$
            \State $e.act \gets get\_next\_activity(\sigma_\text{prefix})$ \label{line:get_activity}
            \State $p\_a \gets \left \{ \right \}$
            \For {$a$ in $A$}
                \If{$e.act$ in $a.c.\textsc{Alloc}$}
                    \State $p\_a \gets add\_to\_potential\_agents(a)$ \label{line:potential_agents}
                \EndIf
            \EndFor
            \State $e.a \gets get\_agent(p\_a)$ \label{line:get_agent}
            \If{$e.a =$ None}
                \State $is\_waiting(\sigma_\text{prefix}) \gets True$ \label{line:waiting}
            \Else{}
                \State $e.ts_{start}, e.ts_{end} \gets execute\_act(e,D)$ \label{line:execute_act}
                \State $\sigma_\text{prefix} \gets add\_to\_prefix(e)$
                \If{$e.act = end$}
                    \State $L', \Sigma \gets exit\_and\_add\_to\_log(\sigma_\text{prefix})$ \label{line:finish_case}
                \EndIf
            \EndIf
        \EndIf
    \EndFor
\EndWhile

\end{algorithmic}
\label{alg:simulation}
\end{algorithm}

\section{Experiments and Results}
\label{sec_evaluation}
\begin{table}[t]
\centering
\setlength\tabcolsep{1.5pt} 
\caption{Description of log properties.}
\label{tab:datasets}
\begin{tabular}{ccrrrrr}
\hline
\toprule
\textbf{Log} & \textbf{Type} & \textbf{\#Traces} & \textbf{\#Events} & \textbf{\#Activities} & \textbf{\#Resources} & \textbf{\#Agents} \\ 
\midrule
 Loan Application & syn  & 1000 & 7492 & 12 & 19 & 19 \\ 
 P2P & syn  & 608 & 9119 & 21 & 27 & 27 \\ 
 CVS & syn  & 10000 & 103906 & 15 & 6 & 8 \\ 
 Confidential 1000 & syn  & 1000 & 38160 & 42 & 14 & 26 \\ 
 Confidential 2000 & syn  & 2000 & 77418 & 42 & 14 & 26 \\ 
 ACR & real  & 954 & 6870 & 18 & 432 & 432 \\ 
 Production & real  & 225 & 4503 & 24 & 41 & 41 \\ 
 BPI12W & real  & 8616 & 59302 & 6 & 52 & 56 \\ 
 BPI17W & real  & 30276 & 240854 & 8 & 136 & 136 \\ 

 \hline
\end{tabular}
\end{table}
This section presents the experiments used to evaluate the performance of AgentSimulator. \autoref{tab:datasets} summarizes the characteristics of the 9 publicly available event logs used in our experiments, which are commonly used in BPS evaluations \cite{camargo_DSIM,meneghello_RIMS,ChapelaCampaBBDKS23} as they contain both start and end timestamps. 
Our implementation, the event logs (with train-test splits), and additional results are available through our public repository\footnote{\url{https://github.com/lukaskirchdorfer/AgentSimulator}}.

\subsection{Experimental Setup}
\label{sec:exp_setup}

\mypar{Implementation} We implemented our approach in Python using the agent-based modeling framework \textit{mesa}~\cite{python-mesa-2020}.

\mypar{Benchmark approaches} We empirically compare our AgentSimulator (AgentSim) against three common data-driven BPS approaches. We adopt the state-of-the-art Data-driven Process Simulation (DDPS) approach from \cite{Lopez-PintadoD22}, which we refer to as Simod. It combines the original Simod \cite{Camargo_2020} with the Prosimos simulator, which considers differentiated resource availability and performance and was shown to outperform the original Simod. DeepGenerator (DGEN)~\cite{camargo_DGEN} is a pure Deep Learning (DL) approach, and DeepSimulator (DSIM)~\cite{camargo_DSIM} is a hybrid of DDPS and DL (more details in \autoref{sec:rel_work}).

\mypar{Data split} We follow evaluations of existing BPS approaches \cite{camargo_DGEN,camargo_DSIM, meneghello_RIMS} and perform a temporal hold-out split, excluding all cases that span the separation time between the train set (first 80\% of cases) and the test set (last 20\% of cases). 

\mypar{Hyperparameters} 
AgentSim has two automatically determined hyperparameters: the architecture (orchestrated or autonomous handovers) and whether to consider extraneous delays, resulting in 4 possible configurations. We treat the latter as a hyperparameter because we noticed considerable differences regarding the benefit of considering extraneous delays between different event logs. We determine both hyperparameters by initially simulating the last 20\% of the training set for each of the 4 possible configurations and checking which simulation most closely resembles the training subset in terms of cycle time. To ensure a fair comparison, we use the same option regarding extraneous delays for Simod.

\mypar{Metrics} To evaluate and compare the different simulation approaches, we use recently proposed metrics that are designed to evaluate simulation models across three dimensions: control-flow, time, and congestion, thus, providing a holistic perspective \cite{ChapelaCampaBBDKS23}. All metrics compute the distance between the simulated and test log, where a lower value indicates a better result. To measure the control-flow, we use the N-Gram Distance (NGD) that computes the difference in the frequencies of the n-grams observed in the event logs. To measure the temporal performance of the simulation, we use the Absolute Event Distribution (AED), the Circadian Event Distribution (CED), and the Relative Event Distribution (RED). To measure the capability of a model to represent congestion, we use the Cycle Time Distribution (CTD). We do not measure the Case Arrival Rate as we apply the same case arrival method as Simod, thus, we focus on metrics where AgentSim differs from other approaches.

\subsection{Results}
\label{sec:results}

\begin{table}[t]
\centering
\setlength\tabcolsep{6pt} 
\caption{Comparison of simulation approaches.}
\label{tab:comparison_all}
\begin{tabular}{cccrrrrr}
\toprule
\multirow{2}{*}{Log} & \multirow{2}{*}{Method} & \multicolumn{6}{c}{Metrics} \\ 
 & & NGD & AED & CED & RED & CTD \\ \midrule
\multirow{4}{*}{\rotatebox[origin=c]{90}{Loan Appl.}} 
 & Simod & \underline{0.15} & \underline{13.55}  & \underline{0.40} & 9.22 & 20.42 \\
 & DGEN & 0.21 & 212.27 & 13.40 & \underline{5.26} & \underline{9.38} \\
 & DSIM & n/a & n/a & n/a & n/a & n/a \\
 & AgentSim & \textbf{0.07} & \textbf{2.78} & \textbf{0.21} & \textbf{1.34} & \textbf{1.49} \\ 
 \midrule
\multirow{4}{*}{\rotatebox[origin=c]{90}{P2P}}
 & Simod & 0.42 & \textbf{1044.25} & 2.21 & 840.19 & 677.05 \\
 & DGEN & \textbf{0.20} & 1481.46 & 2.55 & 828.09 & 670.05 \\
 & DSIM & \underline{0.22} & 1310.03 & \underline{1.15} & \underline{722.33} & \underline{566.63} \\
 & AgentSim & 0.25 & \underline{1161.32} & \textbf{1.02} & \textbf{658.61} & \textbf{525.15} \\
 \midrule
\multirow{4}{*}{\rotatebox[origin=c]{90}{CVS}}
 & Simod & 0.44 & \underline{52.95} & \textbf{0.44} & \underline{39.43} & \underline{54.59} \\
 & DGEN & 0.21 & 310.39 & 11.69 & 176.65 & 294.21 \\
 & DSIM & \underline{0.20} & \textbf{36.23} & 8.98 & \textbf{19.74} & \textbf{52.43} \\
 & AgentSim & \textbf{0.12} & 89.31 & \underline{7.48} & 81.76 & 101.49 \\ 
  \midrule
\multirow{4}{*}{\rotatebox[origin=c]{90}{C. 1000}}
 & Simod & \underline{0.25} & 344.48 & 3.01 & 468.81 & 804.07 \\
 & DGEN & 0.58 & 462.84 & 18.93 & \underline{8.11} & \underline{13.92} \\
 & DSIM & \textbf{0.20} & \underline{246.41} & \underline{2.28} & \textbf{5.34} & \textbf{7.29} \\
 & AgentSim & \underline{0.25} & \textbf{127.01} & \textbf{1.68} & 16.84 & 26.10 \\ 
  \midrule
\multirow{4}{*}{\rotatebox[origin=c]{90}{C. 2000}}
 & Simod & 0.24 & 820.45 & 2.96 & 952.37 & 1614.91 \\
 & DGEN & \textbf{0.16} & 857.68 & 18.09 & \underline{4.58} & \underline{8.12} \\
 & DSIM & \underline{0.18} & \underline{591.13} & \underline{2.84} & \textbf{1.7} & \textbf{2.26} \\
 & AgentSim & 0.26 & \textbf{212.54} & \textbf{1.41} & 9.29 & 17.87 \\ 
  \midrule
\multirow{4}{*}{\rotatebox[origin=c]{90}{ACR}}
 & Simod & \textbf{0.23} & \underline{287.27} & \textbf{2.60} & 32.46 & 93.51 & \\
 & DGEN & 0.31 & 559.67 & 17.84 & 30.87 & 95.11 \\
 & DSIM & \underline{0.26} & \textbf{273.46} & \underline{4.64} & \textbf{15.62} & \textbf{48.24} \\
 & AgentSim & 0.36 & 333.28 & 7.95 & \underline{27.12} & \underline{76.50} \\ 
  \midrule
\multirow{4}{*}{\rotatebox[origin=c]{90}{Production}}
& Simod & 0.93 & \underline{146.38} & \underline{2.82} & 83.88 & 89.15 \\
 & DGEN & \textbf{0.52} & 224.45 & 9.30 & 70.11 & 90.82 \\
 & DSIM & 0.86 & 154.31 & \textbf{2.66} & \underline{33.30} & \underline{43.26} \\
 & AgentSim & \underline{0.61} & \textbf{65.29} & 5.83 & \textbf{14.20} & \textbf{25.79} \\ 
  \midrule
\multirow{4}{*}{\rotatebox[origin=c]{90}{BPI12W}}
 & Simod & 0.72 & \textbf{71.97} & \textbf{1.71} & \underline{95.72} & \underline{155.46} \\
 & DGEN & \underline{0.43} & 306.28 & 4.53 & 116.18 & 176.79 \\
 & DSIM & 0.65 & \underline{78.62} & 2.88 & 119.12 & 173.49 \\
 & AgentSim & \textbf{0.15} & 79.89 & \underline{1.87} & \textbf{47.83} & \textbf{90.12} \\ 
  \midrule
\multirow{4}{*}{\rotatebox[origin=c]{90}{BPI17W}}
 & Simod & 0.59 & 300.28 & \underline{3.34} & 136.63 & 148.40 \\
 & DGEN & 0.67 & 4557.19 & 3.39 & 118.84 & 172.94 \\
 & DSIM & \underline{0.53} & \textbf{54.61} & 3.35 & \underline{33.10} & \underline{30.26} \\
 & AgentSim & \textbf{0.30} & \underline{220.98} & \textbf{1.64} & \textbf{26.03} & \textbf{22.75} \\ 
 \bottomrule
\end{tabular}
\end{table}

\mypar{Overall results} \autoref{tab:comparison_all} summarizes the results for the 9 event logs, showing the average metrics from 10 simulation runs for each log. The best value per log and metric is marked in bold, the second-best is underlined. Generally, no single approach consistently outperforms the others across all datasets and metrics as also observed in previous BPS works \cite{meneghello_RIMS,camargo_DSIM}. However, AgentSim stands out by most frequently achieving the best performance within each metric across the 9 logs. When looking more closely at the 3 dimensions captured by the metrics, we can obtain the following main insights:

\mypartwo{Control-flow} Achieving the best NGD (N-gram distance) in 4 out of 9 logs and considerably outperforming Simod and DSIM, the results indicate that our resource-first AgentSim approach, which is independent of an underlying process model enhances the accuracy of the control-flow dimension compared to control-flow-first approaches such as Simod. This is particularly evident when analyzing the two real-life BPI logs. Thus, putting the resource at the core of the simulation often allows to represent control-flow behavior more faithfully.

\mypartwo{Time} In terms of absolute (AED), circadian (CED), and relative (RED) event distributions over time, AgentSim is the leader across all 3 metrics, followed by DSIM. Simod and DGEN stay on average significantly behind AgentSim. These results indicate that AgentSim can capture the temporal patterns in the logcomparatively well.

\mypartwo{Congestion} Accurately capturing the cycle time of a process instance serves as a critical indicator of the accuracy of a simulation approach. The CTD metric is influenced by the processing times of individual activities and the corresponding waiting times, which in turn are dependent on resource availability. Consequently, the cycle time metric captures a comprehensive range of factors. AgentSim demonstrates superior accuracy in capturing cycle times compared to the benchmarks, achieving the best CTD in 5 out of 9 logs, again followed by DSIM leading in the remaining 4 logs. The advantage of AgentSim over Simod is particularly pronounced with the Confidential logs. Here, Simod's simulated logs exhibit significantly longer cycle times compared to reality, primarily due to resource contention and the consequent waiting times. AgentSim mitigates this issue by recognizing that some agents do not require waiting time, such as agents performing instantaneous activities.

\mypar{Impact of handover configuration} Simulating the same process using the two different handover configurations (centrally orchestrated versus autonomous) reveals considerable differences in results for some logs (full results for both options are in our repository). Overall, our automated method of selecting the handover configuration in AgentSim leads to 5 of 9 processes being simulated with orchestrated handovers, e.g., P2P and Production, for which AgentSim, despite being a resource-first approach, still yields strong results. 
Some logs strongly benefit from being simulated in an orchestrated instead of autonomous manner. For instance, the Production log shows an improvement in CTD from 45.65 to 25.79 and in NGD from 0.77 to 0.61, aligning with the standardized nature of production processes. 
By contrast, looking at the logs that are simulated with autonomous handovers, we also observe clear benefits of considering distinct agent behaviors for certain processes. For instance, for the real-life BPI17W log, we halve both the CTD and RED to 22.75 and 26.03, respectively, compared to the orchestrated configuration. Also, C. 1000 and Loan Appl. show slightly better results across all metrics using autonomous handovers. Overall, these insights demonstrate AgentSim's adaptability to diverse process types, leading to considerable improvements in terms of results.

\mypar{Post-hoc analysis on agent interactions} To exemplarily demonstrate AgentSim's ability to represent agent interactions, we compare its interaction patterns with Simod for the real-life BPI12W log. \autoref{fig:agent_interactions} shows that AgentSim accurately reflects the training log's interaction dynamics, capturing the given activity chaining of resources (see diagonal). In contrast, Simod exhibits random interactions and fails to capture these patterns, thus showing the crucial nature of considering individual interaction patterns for certain processes.

\begin{figure}[t]
\vspace{-1em}
    \centering
    \includegraphics[trim = 35mm 0mm 35mm 0mm, clip, width=\columnwidth]{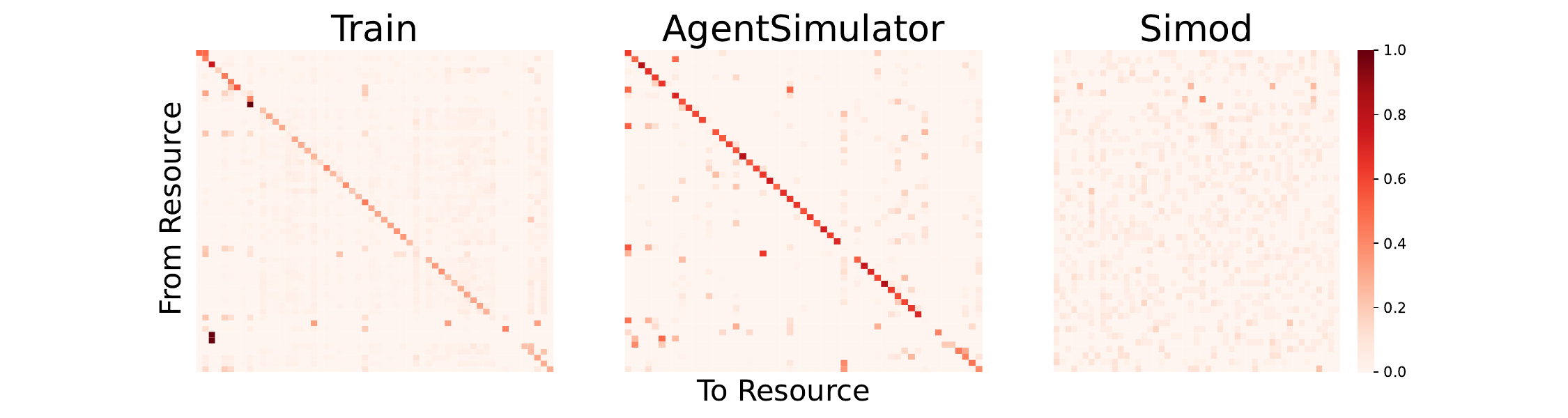}
    \caption{Resource interactions for the BPI12W log.}
    \label{fig:agent_interactions}
\end{figure}

\mypar{Runtime}
Given the stochasticity of simulation, it is essential to simulate numerous logs to achieve reliable predictions. Therefore, runtime is a critical factor in the practical application of BPS. 
In this regard, it is important to note that AgentSim consistently operates much faster than the benchmarks. 
For instance, while AgentSim only requires around 30 seconds to discover and simulate the Production log (on a machine with 32GB of RAM and an Intel Core i7 2.3 GHz CPU), Simod runs 20 times and DSIM even 80 times longer. For the BPI12W log, AgentSim requires around 9 minutes, whereas DSIM takes more than 10 hours.

\section{Related Work}
\label{sec:rel_work}
This section briefly discusses related work on automated BPS and agent-based modeling and simulation.

\mypar{Automated BPS} 
We can divide existing literature on automated BPS approaches into three categories: Data-Driven Process Simulation (DDPS), Deep Learning (DL), and hybrid approaches. DDPS approaches automate simulation model discovery from event logs by initially identifying a process model and then enhancing it with simulation parameters. A semi-automated approach using colored Petri nets is proposed in~\cite{RozinatMSA09}, while~\cite{KhodyrevP14} introduces a data-driven approach without considering resources. 
A more recent approach is Simod~\cite{Camargo_2020}, which incorporates hyperparameter tuning. 
DL approaches for BPS typically rely on recurrent neural networks. LSTM models are employed in~\cite{TaxVRD17} to predict events and timestamps, later improved upon by DGEN~\cite{camargo_DGEN} incorporating n-grams and embeddings. Due to their black-box nature, DL models are not applicable for what-if analysis.
Hybrid models combine DDPS and DL approaches. DSIM~\cite{camargo_DSIM} combines a stochastic process model with DL for event timestamping, extended by RIMS that integrates predictions at runtime~\cite{meneghello_RIMS}.

\mypar{Agent-based modeling and simulation} Over the past decades, the application of MAS to various domains has been studied extensively (cf.~\cite{DorriKJ18} for a review). Applying agents to Business Process Management (BPM) was initially proposed in the 1990's~\cite{JenningsFJNOW96}, where a business process is modeled as a system of negotiating agents. More recently, the concept of agent system mining has been introduced, recognizing that processes often emerge from interactions of autonomous agents~\cite{TourPK21}, as demonstrated by an agent-based discovery algorithm in~\cite{TourPKS23}, and also shown for simulation in~\cite{HalaškaŠperka+2018+255+269}. A general introduction to agent-based BPS can be found in~\cite{SulisT22}. However, such agent-based simulation approaches in BPM rely on manual configurations to simulate a specific process, e.g., in a factory production domain~\cite{jmmp4030089}. To the best of our knowledge, our approach is the first to use event logs to automatically infer MAS models for process simulation.

\section{Conclusion}
\label{sec:conclusion}
This paper introduced AgentSimulator---an agent-based approach for data-driven business process simulation. Given an event log, our approach discovers a multi-agent system that represents real-world actors and systems, each modeled with unique behaviors and interaction patterns. The discovered multi-agent system is then used to simulate the execution of the process. Our resource-first approach provides more means to capture distinct resource behaviors and interactions than traditional control-flow-first approaches and achieves state-of-the-art results with significantly reduced computation times. The evaluation shows that centrally orchestrated and decentralized processes often need to be captured differently, with AgentSimulator being automatically adaptable to both. Modeling human behavior is a complex task. Although our approach successfully captures some agent-specific behaviors, it currently does not account for multitasking, batching, or fatigue effects, which we want to incorporate in future work. Furthermore, we will explore the use of additional architectures for our multi-agent system itself.

\bibliographystyle{IEEEtran}
\bibliography{main}

\begin{thebibliography}{10}
\providecommand{\url}[1]{#1}
\csname url@samestyle\endcsname
\providecommand{\newblock}{\relax}
\providecommand{\bibinfo}[2]{#2}
\providecommand{\BIBentrySTDinterwordspacing}{\spaceskip=0pt\relax}
\providecommand{\BIBentryALTinterwordstretchfactor}{4}
\providecommand{\BIBentryALTinterwordspacing}{\spaceskip=\fontdimen2\font plus
\BIBentryALTinterwordstretchfactor\fontdimen3\font minus \fontdimen4\font\relax}
\providecommand{\BIBforeignlanguage}[2]{{%
\expandafter\ifx\csname l@#1\endcsname\relax
\typeout{** WARNING: IEEEtran.bst: No hyphenation pattern has been}%
\typeout{** loaded for the language `#1'. Using the pattern for}%
\typeout{** the default language instead.}%
\else
\language=\csname l@#1\endcsname
\fi
#2}}
\providecommand{\BIBdecl}{\relax}
\BIBdecl

\bibitem{FundamentalsOfBPM}
M.~Dumas, M.~L. Rosa, J.~Mendling, and H.~A. Reijers, \emph{Fundamentals of Business Process Management}.\hskip 1em plus 0.5em minus 0.4em\relax Springer, 2013.

\bibitem{Aalst15}
W.~M.~P. van~der Aalst, ``Business process simulation survival guide,'' in \emph{Handbook on Business Process Management 1, Introduction, Methods, and Information Systems, 2nd Ed}, ser. International Handbooks on Information Systems, J.~vom Brocke and M.~Rosemann, Eds.\hskip 1em plus 0.5em minus 0.4em\relax Springer, 2015, pp. 337--370.

\bibitem{RozinatMSA09}
A.~Rozinat, R.~S. Mans, M.~Song, and W.~M.~P. van~der Aalst, ``Discovering simulation models,'' \emph{Inf. Syst.}, vol.~34, no.~3, pp. 305--327, 2009.

\bibitem{Camargo_2020}
M.~Camargo, M.~Dumas, and O.~González-Rojas, ``Automated discovery of business process simulation models from event logs,'' \emph{Decision Support Systems}, vol. 134, p. 113284, 2020.

\bibitem{meneghello_RIMS}
F.~Meneghello, C.~D. Francescomarino, and C.~Ghidini, ``Runtime integration of machine learning and simulation for business processes,'' in \emph{ICPM}.\hskip 1em plus 0.5em minus 0.4em\relax {IEEE}, 2023.

\bibitem{camargo_DSIM}
M.~Camargo, M.~Dumas, and O.~G. Rojas, ``Learning accurate business process simulation models from event logs via automated process discovery and deep learning,'' in \emph{CAiSE}.\hskip 1em plus 0.5em minus 0.4em\relax Springer, 2022.

\bibitem{KhodyrevP14}
I.~Khodyrev and S.~Popova, ``Discrete modeling and simulation of business processes using event logs,'' in \emph{ICCS}.\hskip 1em plus 0.5em minus 0.4em\relax Elsevier, 2014.

\bibitem{Lopez-PintadoD22}
O.~L{\'{o}}pez{-}Pintado and M.~Dumas, ``Business process simulation with differentiated resources: Does it make a difference?'' in \emph{BPM}.\hskip 1em plus 0.5em minus 0.4em\relax Springer, 2022.

\bibitem{RN2020}
S.~Russell and P.~Norvig, \emph{Artificial Intelligence: {A} Modern Approach (4th Edition)}.\hskip 1em plus 0.5em minus 0.4em\relax Pearson, 2020.

\bibitem{Song2008}
M.~Song and W.~Aalst, ``Towards comprehensive support for organizational mining,'' \emph{Decision Support Systems}, vol.~46, pp. 300--317, 2008.

\bibitem{LopezPintado_Prob_Calendars}
O.~L{\'{o}}pez{-}Pintado and M.~Dumas, ``Discovery and simulation of business processes with probabilistic resource availability calendars,'' in \emph{ICPM}.\hskip 1em plus 0.5em minus 0.4em\relax {IEEE}, 2023.

\bibitem{camargo_DGEN}
M.~Camargo, M.~Dumas, and O.~G. Rojas, ``Learning accurate {LSTM} models of business processes,'' in \emph{BPM}.\hskip 1em plus 0.5em minus 0.4em\relax Springer, 2019.

\bibitem{TaxVRD17}
N.~Tax, I.~Verenich, M.~L. Rosa, and M.~Dumas, ``Predictive business process monitoring with {LSTM} neural networks,'' in \emph{CAiSE}.\hskip 1em plus 0.5em minus 0.4em\relax Springer, 2017.

\bibitem{chapela2024enhancing}
D.~Chapela-Campa and M.~Dumas, ``Enhancing business process simulation models with extraneous activity delays,'' \emph{Information Systems}, vol. 122, p. 102346, 2024.

\bibitem{ChapelaCampaBBDKS23}
D.~Chapela{-}Campa, I.~Benchekroun, O.~Baron, M.~Dumas, D.~Krass, and A.~Senderovich, ``Can {I} trust my simulation model? measuring the quality of business process simulation models,'' in \emph{BPM}.\hskip 1em plus 0.5em minus 0.4em\relax Springer, 2023.

\bibitem{python-mesa-2020}
J.~Kazil, D.~Masad, and A.~T. Crooks, ``Utilizing python for agent-based modeling: The mesa framework,'' vol. 12268.\hskip 1em plus 0.5em minus 0.4em\relax Springer, 2020, pp. 308--317.

\bibitem{DorriKJ18}
A.~Dorri, S.~S. Kanhere, and R.~Jurdak, ``Multi-agent systems: {A} survey,'' \emph{{IEEE} Access}, vol.~6, pp. 28\,573--28\,593, 2018.

\bibitem{JenningsFJNOW96}
N.~R. Jennings, P.~Faratin, M.~J. Johnson, T.~J. Norman, P.~D. O'Brien, and M.~E. Wiegand, ``Agent-based business process management,'' \emph{Int. J. Cooperative Inf. Syst.}, vol.~5, no. 2{\&}3, pp. 105--130, 1996.

\bibitem{TourPK21}
A.~Tour, A.~Polyvyanyy, and A.~A. Kalenkova, ``Agent system mining: Vision, benefits, and challenges,'' \emph{{IEEE} Access}, vol.~9, pp. 99\,480--99\,494, 2021.

\bibitem{TourPKS23}
A.~Tour, A.~Polyvyanyy, A.~A. Kalenkova, and A.~Senderovich, ``Agent miner: An algorithm for discovering agent systems from event data,'' in \emph{BPM}.\hskip 1em plus 0.5em minus 0.4em\relax Springer, 2023.

\bibitem{HalaškaŠperka+2018+255+269}
M.~Halaška and R.~Šperka, ``Is there a need for agent-based modelling and simulation in business process management?'' \emph{Organizacija}, vol.~51, no.~4, pp. 255--269, 2018.

\bibitem{SulisT22}
E.~Sulis and K.~Taveter, \emph{Agent-Based Business Process Simulation - {A} Primer with Applications and Examples}.\hskip 1em plus 0.5em minus 0.4em\relax Springer, 2022.

\bibitem{jmmp4030089}
M.~Dornhöfer, S.~Sack, J.~Zenkert, and M.~Fathi, ``Simulation of smart factory processes applying multi-agent-systems—a knowledge management perspective,'' \emph{JMMP}, vol.~4, no.~3, 2020.

\end{thebibliography}

\end{document}